\documentclass[manyauthors,nocleardouble,COMPASS]{cernphprep}
\usepackage{bm} % blackboard bold                                         %
\usepackage{cite}                                                         %
\usepackage[figuresright]{rotating}
\newcommand{\id}{\mathrm{d}}
\newcommand{\GeV}{\mathrm{GeV}}
\begin {document}

\begin{titlepage}
\PHnumber{2010--023}
\PHdate{21 July 2010}

\title{
 Quark helicity distributions from longitudinal
 spin asymmetries in muon--proton and muon--deuteron scattering
}  

\Collaboration{The COMPASS Collaboration}
\ShortAuthor{The COMPASS Collaboration}
\ShortTitle{Quark helicity distributions \ldots}

\begin{abstract}
Double-spin asymmetries for production of charged pions and kaons
in semi-inclusive deep-inelastic muon scattering have been
measured by the COMPASS experiment at CERN. The data, obtained by
scattering a 160~GeV muon beam off a longitudinally polarised
NH$_3$ target, cover a range of the Bjorken variable $x$ between
0.004 and 0.7. A
leading order evaluation of the  helicity distributions for the
three lightest quarks and antiquark flavours
derived from these asymmetries and from our previous deuteron data
is presented. The resulting values of the sea quark distributions
are small and  do not show any sizable
dependence on $x$ in the range of the measurements.
No significant difference is observed between the strange
%, $\Delta s$, and antistrange, $\Delta {\overline s}$,
and antistrange helicity distributions, both compatible with
zero.
The integrated value of the flavour asymmetry of the helicity
distribution of the light-quark sea,
$\Delta {\overline u} - \Delta {\overline d}$,
is found to be slightly positive, about $1.5$ standard
deviations away from zero.
\end{abstract}

\vspace*{60pt}
\begin{flushleft}
%PACS:   {13.60.Hb}, % Total and inclusive cross sections
%        {13.85.Hd}, % Inelastic scattering, many particle final states 
%        {13.85.Ni}, % Inclusive production with identified hadrons 
%        {13.88.+e}\\  % Polarization in interaction and scattering
\end{flushleft}

\vfill
\Submitted{(Submitted to Physics Letters B)}
\end{titlepage}

{\pagestyle{empty}
%%%%%%%%%%%%%%%%%%%%%%%%%%%%%%%%%%%%%%%%%%%%%%%%%%%%%%%%%%%%%%%%%%%%%%%%%%%%%%%%%%%%%%%%%%%%%%%%%%%%%%%%%%%%%%%%%%%%%%%
%
% 2010_longitudity/auth_cern.tex  
%
%%%%%%%%%%%%%%%%%%%%%%%%%%%%%%%%%%%%%%%%%%%%%%%%%%%%%%%%%%%%%%%%%%%%%%%%%%%%%%%%%%%%%%%%%%%%%%%%%%%%%%%%%%%%%%%%%%%%%%%

\section*{The COMPASS Collaboration}
\label{app:collab}

\begin{flushleft}
M.G.~Alekseev\Irefn{turin_i},
V.Yu.~Alexakhin\Irefn{dubna},
Yu.~Alexandrov\Irefn{moscowlpi},
G.D.~Alexeev\Irefn{dubna},
A.~Amoroso\Irefn{turin_u},
A.~Austregesilo\Irefnn{cern}{munichtu},
B.~Bade{\l}ek\Irefn{warsaw},
F.~Balestra\Irefn{turin_u},
%J.~Ball\Irefn{saclay},
J.~Barth\Irefn{bonnpi},
G.~Baum\Irefn{bielefeld}
% ...
\end{flushleft}

%%%%%%%%%%%%%%%%%%%%%%%%%%%%%%%%%%%%%%%%%%%%%%%%%%%%%%%%%%%%%%%%%%%%%%%%%%%%%%%%%%%%%%%%%%%%%%%%%%%%%%%%%%%%%%%%%%%%%%%
%
% institutes
%
%%%%%%%%%%%%%%%%%%%%%%%%%%%%%%%%%%%%%%%%%%%%%%%%%%%%%%%%%%%%%%%%%%%%%%%%%%%%%%%%%%%%%%%%%%%%%%%%%%%%%%%%%%%%%%%%%%%%%%%

\begin{Authlist}
\item \Idef{bielefeld}{Universit\"at Bielefeld, Fakult\"at f\"ur Physik, 33501 Bielefeld, Germany\Arefs{f}}
\item \Idef{bochum}{Universit\"at Bochum, Institut f\"ur Experimentalphysik, 44780 Bochum, Germany\Arefs{f}}
\item \Idef{bonniskp}{Universit\"at Bonn, Helmholtz-Institut f\"ur  Strahlen- und Kernphysik, 53115 Bonn, Germany\Arefs{f}}
\item \Idef{bonnpi}{Universit\"at Bonn, Physikalisches Institut, 53115 Bonn, Germany\Arefs{f}}
\item \Idef{brno}{Institute of Scientific Instruments, AS CR, 61264 Brno, Czech Republic\Arefs{g}}
%\item \Idef{burdwan}{Burdwan University, Burdwan 713104, India\Arefs{h}}
\item \Idef{calcutta}{Matrivani Institute of Experimental Research \& Education, Calcutta-700 030, India\Arefs{h}}
\item \Idef{dubna}{Joint Institute for Nuclear Research, 141980 Dubna, Moscow region, Russia\Arefs{i}}
\item \Idef{erlangen}{Universit\"at Erlangen--N\"urnberg, Physikalisches Institut, 91054 Erlangen, Germany\Arefs{f}}
\item \Idef{freiburg}{Universit\"at Freiburg, Physikalisches Institut, 79104 Freiburg, Germany\Arefs{f}}
\item \Idef{cern}{CERN, 1211 Geneva 23, Switzerland}
%\item \Idef{heidelberg}{Universit\"at Heidelberg, Physikalisches Institut,  69120 Heidelberg, Germany\Arefs{e}}
%\item \Idef{helsinki}{Helsinki University of Technology, Low Temperature Laboratory, 02015 HUT, Finland  and University of Helsinki, Helsinki Institute of  Physics, 00014 Helsinki, Finland}
\item \Idef{liberec}{Technical University in Liberec, 46117 Liberec, Czech Republic\Arefs{g}}
\item \Idef{lisbon}{LIP, 1000-149 Lisbon, Portugal\Arefs{j}}
\item \Idef{mainz}{Universit\"at Mainz, Institut f\"ur Kernphysik, 55099 Mainz, Germany\Arefs{f}}
\item \Idef{miyazaki}{University of Miyazaki, Miyazaki 889-2192, Japan\Arefs{k}}
\item \Idef{moscowlpi}{Lebedev Physical Institute, 119991 Moscow, Russia}
\item \Idef{munichlmu}{Ludwig-Maximilians-Universit\"at M\"unchen, Department f\"ur Physik, 80799 Munich, Germany\AAref{f}{l}}
\item \Idef{munichtu}{Technische Universit\"at M\"unchen, Physik Department, 85748 Garching, Germany\AAref{f}{l}}
%\item \Idef{munichtucl}{Excellence Cluster Universe, Technische Universit\"at M\"unchen, Physik Department, 85748 Garching, Germany\AAref{f}{l}}
\item \Idef{nagoya}{Nagoya University, 464 Nagoya, Japan\Arefs{k}}
\item \Idef{praguecu}{Charles University in Prague, Faculty of Mathematics and Physics, 18000 Prague, Czech Republic\Arefs{g}}
\item \Idef{praguectu}{Czech Technical University in Prague, 16636 Prague, Czech Republic\Arefs{g}}
\item \Idef{protvino}{State Research Center of the Russian Federation, Institute for High Energy Physics, 142281 Protvino, Russia}
\item \Idef{saclay}{CEA IRFU/SPhN Saclay, 91191 Gif-sur-Yvette, France}
\item \Idef{telaviv}{Tel Aviv University, School of Physics and Astronomy, 69978 Tel Aviv, Israel\Arefs{m}}
\item \Idef{triest_i}{Trieste Section of INFN, 34127 Trieste, Italy}
\item \Idef{triest}{University of Trieste, Department of Physics and Trieste Section of INFN, 34127 Trieste, Italy}
\item \Idef{triestictp}{Abdus Salam ICTP and Trieste Section of INFN, 34127 Trieste, Italy}
\item \Idef{turin_u}{University of Turin, Department of Physics and Torino Section of INFN, 10125 Turin, Italy}
\item \Idef{turin_i}{Torino Section of INFN, 10125 Turin, Italy}
\item \Idef{turin_p}{University of Eastern Piedmont, 1500 Alessandria,  and Torino Section of INFN, 10125 Turin, Italy}
\item \Idef{warsaw}{So{\l}tan Institute for Nuclear Studies and University of Warsaw, 00-681 Warsaw, Poland\Arefs{n} }
\item \Idef{warsawtu}{Warsaw University of Technology, Institute of Radioelectronics, 00-665 Warsaw, Poland\Arefs{n} }
\item \Idef{yamagata}{Yamagata University, Yamagata, 992-8510 Japan\Arefs{k} }

\end{Authlist}

%%%%%%%%%%%%%%%%%%%%%%%%%%%%%%%%%%%%%%%%%%%%%%%%%%%%%%%%%%%%%%%%%%%%%%%%%%%%%%%%%%%%%%%%%%%%%%%%%%%%%%%%%%%%%%%%%%%%%%%
%
% Notes
%
%%%%%%%%%%%%%%%%%%%%%%%%%%%%%%%%%%%%%%%%%%%%%%%%%%%%%%%%%%%%%%%%%%%%%%%%%%%%%%%%%%%%%%%%%%%%%%%%%%%%%%%%%%%%%%%%%%%%%%%
\vspace*{-\baselineskip}
\begin{Authlist}
\item \Adef{a}{Also at IST, Universidade T\'ecnica de Lisboa, Lisbon, Portugal}
\item \Adef{b}{Also at Chubu University, Kasugai, Aichi, 487-8501 Japan\Arefs{k}}
\item \Adef{c}{Also at KEK, 1-1 Oho, Tsukuba, Ibaraki, 305-0801 Japan}
\item \Adef{d}{On leave of absence from JINR Dubna}
\item \Adef{e}{Also at GSI mbH, Planckstr.\ 1, D-64291 Darmstadt, Germany}
\item \Adef{f}{Supported by the German Bundesministerium f\"ur Bildung und Forschung}
\item \Adef{g}{Suppported by Czech Republic MEYS grants ME492 and LA242}
\item \Adef{h}{Supported by SAIL (CSR), Govt.\ of India}
%
% transversity only CERN-RFBR grant 08-02-91009
%
%\item \Adef{i}{Supported by CERN-RFBR grants 08-02-91009 and 08-02-91013}
\item \Adef{i}{Supported by CERN-RFBR grants 08-02-91009}
\item \Adef{j}{\raggedright Supported by the Portuguese FCT - Funda\c{c}\~{a}o para a
             Ci\^{e}ncia e Tecnologia grants POCTI/FNU/49501/2002 and POCTI/FNU/50192/2003}
\item \Adef{k}{Supported by the MEXT and the JSPS under the Grants No.18002006, No.20540299 and No.18540281; Daiko Foundation and Yamada Foundation}
\item \Adef{l}{Supported by the DFG cluster of excellence `Origin and Structure of the Universe' (www.universe-cluster.de)}
\item \Adef{m}{Supported by the Israel Science Foundation, founded by the Israel Academy of Sciences and Humanities}
\item \Adef{n}{Supported by Ministry of Science and Higher Education grant 41/N-CERN/2007/0}
%\item \Adef{o}{Supported by KBN grant nr 134/E-365/SPUB-M/CERN/P-03/DZ299/2000}
\item [{\makebox[2mm][l]{\textsuperscript{*}}}] Deceased
\end{Authlist}
    
\clearpage
}
%=============================  

\setcounter{page}{1}

\section {Introduction}

Measurements of spin asymmetries in polarised Deep 
Inelastic Scattering (DIS) provide the main source of
information on the spin structure of the nucleon. 
Semi-inclusive DIS (SIDIS) cross-section asymmetries,
where in addition to the scattered lepton, pions and 
kaons are  detected, are sensitive to the individual 
quark and antiquark flavours. 
Together with DIS and polarised proton--proton data, they 
are a key ingredient in QCD fits aiming at the evaluation 
of the quark helicity distributions\cite{dssv}. Although 
evaluations of the parton helicity distributions do exist, 
their uncertainties are still important, particularly for 
low values of the    
Bjorken scaling variable $x$.  
Data with improved precision, covering a large region of 
$x$, should help to clarify several challenging  observations 
made in the last few years. 
The first one is the flavour asymmetry of the light-quark 
sea. Unpolarised lepton scattering 
\cite{Amau91,Arneo94,Acke98}, later confirmed by Drell--Yan 
production experiments\cite{Bald94,Towe01}, has revealed 
that the light-quark difference, ${\overline u}-{\overline d}$, 
is negative and much larger in absolute value 
than expected, resulting in a large violation of the 
Gottfried sum rule\cite{GP01}. Model calculations that explain 
this asymmetry also provide 
predictions\cite{bhalerao,peng,bsb} for the polarised 
flavour asymmetry, $\Delta {\overline u} - \Delta {\overline d}$.

The total contribution of the strange quarks, $\Delta s$, 
to the nucleon spin is presently another intriguing issue. 
Assuming SU(3)$_{\rm f}$ flavour symmetry, global fits \cite{LSS07,ACC06} 
on inclusive DIS data favour a large and negative first moment 
for the strange contribution, $\Delta s + \Delta {\overline s} 
\approx -0.10$. Combined analysis of parity violating $\vec{e}p$ 
asymmetries and $\nu p$ elastic data, used for the 
extraction of the strange axial form factor, also suggests a 
negative strange quark contribution \cite{Pate}. 
Surprisingly, these results are at variance with 
the most recent determinations of the strange quark 
distribution $\Delta s(x)$  from
% Semi-Inclusive Deep Inelastic Scattering 
SIDIS,  which appears to be compatible with zero \cite{hrm1} 
or even slightly positive\cite{hrm8}, at least in the $x$ 
range of the measurements. 
Among possible explanations for this observation, a 
substantial breaking of the SU(3)$_{\rm f}$ symmetry has been 
advocated \cite{bass}. 
The distribution of $\Delta s(x)$  may also change sign at 
lower values of $x$. This option is used in a global fit to 
DIS, SIDIS and proton--proton data \cite{dssv} to reconcile 
the medium-$x$ SIDIS data with a large and negative first moment. 
The dependence of the results on the Fragmentation Functions (FF) 
and particularly on the strange-quark FF may also introduce a 
significant bias\cite{flavsep} and for this reason must be 
quantified.
Finally, phenomenological studies usually assume that the 
strange and antistrange quark helicity distributions are 
identical, an assumption which has never been tested 
experimentally. For these reasons a determination of both 
the shape and the magnitude of the flavour-separated quark 
helicity distributions is necessary, particularly towards 
lower values of $x$.
 
The first measurements of SIDIS  asymmetries were performed by 
the EMC 
Collaboration more
than twenty years ago \cite{emcsidis}.
More recently, the SMC Collaboration has measured SIDIS 
asymmetries for unidentified charged hadrons \cite{smc}. 
The HERMES Collaboration has reported 
SIDIS asymmetries for charged pion production on a proton target 
and for charged pion
and kaon production on a deuteron target
\cite{hrm1}. These asymmetries were used for a flavour 
decomposition into 
five helicity distributions. However the data did not permit 
an extraction of
$\Delta {\overline s}$.
In a previous publication, we have presented a leading-order 
(LO) 
evaluation of the isoscalar polarised valence, sea and 
strange distributions, $\Delta u_v + \Delta d_v$, 
$\Delta {\overline u} + \Delta {\overline d}$ and 
$(\Delta s + \Delta {\overline s})/2$,
all  derived from DIS
and SIDIS asymmetries on a polarised deuteron target 
only\cite{flavsep}.

In this Letter we present new semi-inclusive asymmetries 
for scattering of high-energy muons off a polarised proton 
target for production of identified charged pions 
($A_{1,p}^{\pi+}$,  $A_{1,p}^{\pi-}$) and, for the first 
time, of identified charged kaons ($A_{1,p}^{K+}$,  
$A_{1,p}^{K-}$). 
To the current measurements we have added our previous 
($A_{1,d}^{\pi+}$, $A_{1,d}^{\pi-}$) and ($A_{1,d}^{K+}$, 
$A_{1,d}^{K-}$) SIDIS deuteron data as well as our inclusive 
double-spin asymmetries $A_{1,p}$ \cite{g1p} and 
$A_{1,d}$\cite{flavsep}. Using these measurements we perform a 
full flavour decomposition in LO, thus accessing 
for the first time 
all up, down and strange quark and antiquark helicity 
distributions separately.

\section {Data}
The data presented below were collected in 2007 using the 
COMPASS spectrometer \cite{spec} at CERN.  A 160~GeV muon beam 
was scattered off a polarised NH$_3$ (ammonia) target. 
The target consists of three consecutive cells with a length of 
30~cm, 60~cm and 30~cm respectively, polarised by Dynamic 
Nuclear Polarisation (DNP).
Neighbouring target cells were always polarised in 
opposite directions.  
The spin orientations were reversed once per day
by rotating the solenoidal magnetic field and, in addition, 
every few weeks by interchanging
the DNP microwave frequencies. 
The beam and target polarisations of  $-0.80$ and about 
$\pm 0.90$
are known with relative precisions of 5\% and 2\%, 
respectively.
The energy of the incident muons was constrained to be in 
the interval
$140~\GeV < E_{\mu} < 180~\GeV$.
The kinematic region was defined by cuts on the photon
virtuality, $Q^2 > 1~(\GeV/c)^2$, and on the fractional 
energy, $y$,
transfered from the incident muon to the virtual photon 
$0.1 < y < 0.9$. The selected data sample covers the range 
$0.004 < x < 0.7$ and
consists of 85.3 million events.

The reconstruction procedure is described in 
Ref.~\cite{rec_proc}.
All events were required to have a reconstructed primary 
vertex inside one of the target cells. 
Only hadron tracks originating from the primary vertex 
were  considered. Their fractional
energy, $z$, was required to be larger than 0.2 in order to 
select hadrons produced 
in the current fragmentation region and smaller than 0.85 in 
order to suppress hadrons originating from diffractive processes. 
Tracks crossing more than 30 radiation lengths were not 
accepted as hadrons. 
Hadrons were identified in the RICH detector. The momentum range 
was restricted to the interval $ 10~\GeV/c < p < 50~\GeV/c$ where 
both pions and kaons can be identified. The measured RICH 
photoelectron distributions were compared to 
parameterisations calculated using the pion and kaon masses, 
taking into account the observed background. The masses of 
the detected hadrons were then assigned according to their 
likelihood ratios. The total samples of $\pi^+$, ($\pi^-$) 
and $K^+$ ($K^-$) amount to
12.3 (10.9) and 3.6 (2.3) million hadrons, respectively.

Since the RICH detector does  not select pure samples of pions 
and kaons, a correction
accounting for the fraction of misidentified hadrons in each
sample was applied.
The  unfolding procedure is the same as that used  for the 
$^6$LiD data in Ref.~\cite{flavsep}. The purity of
the pion samples selected by the RICH detector 
is larger than 0.98 over the full range of $x$, 
while for kaons it varies from about 0.73 at the lowest value 
of $x$ to about 0.93 and 0.91 at $x \ge 0.03$ for positive and 
negative kaons, respectively. The applied unfolding corrections 
have only a small effect on the pion and kaon asymmetries.

\section {Asymmetries}

Spin asymmetries were calculated from the numbers of hadrons 
originating
from cells with opposite spin orientations collected before and
after a target field rotation so that flux and acceptance factors 
cancel out.
Corrections were applied for QED radiative effects \cite{radcor} 
and  for the polarisation
of the $^{14}$N nucleus. The latter is proportional to the 
corresponding 
asymmetry for a deuteron target and approximately given by 
\cite{rondon}:
\begin{equation}
\Delta A_{1,p}^h = \frac{1}{3} \cdot (-\frac{1}{3}) 
\cdot \frac{1}{6} \cdot
\frac{\sigma_d^h(x)}{\sigma_p^h(x)} \cdot A_{1,d}^h(x)
\end{equation}
where $h = \pi^+,\pi^-,K^+,K^-$.
The factors account for the fraction of polarisable nitrogen 
nucleons in ammonia, the
alignment of the proton spin vs.\ the $^{14}$N spin, the ratio 
of $^{14}$N to
$^1$H polarisations and the ratio of the hadron cross sections, 
$\sigma_d^h$ and $\sigma_p^h$,  for scattering of muons off  
unpolarised deuteron and proton targets.
The correction thus varies both with $x$ and the type of hadron:
for $\pi^+$, $\pi^-$ and $K^{+}$ it reaches about $-0.015$ at 
$x = 0.5$, while for 
$K^{-}$
it remains practically equal to zero due to the vanishing values
of $A_{1,d}^{K-}$.

\begin{figure}[tb]
\centering
\includegraphics[width=0.95\textwidth,clip]{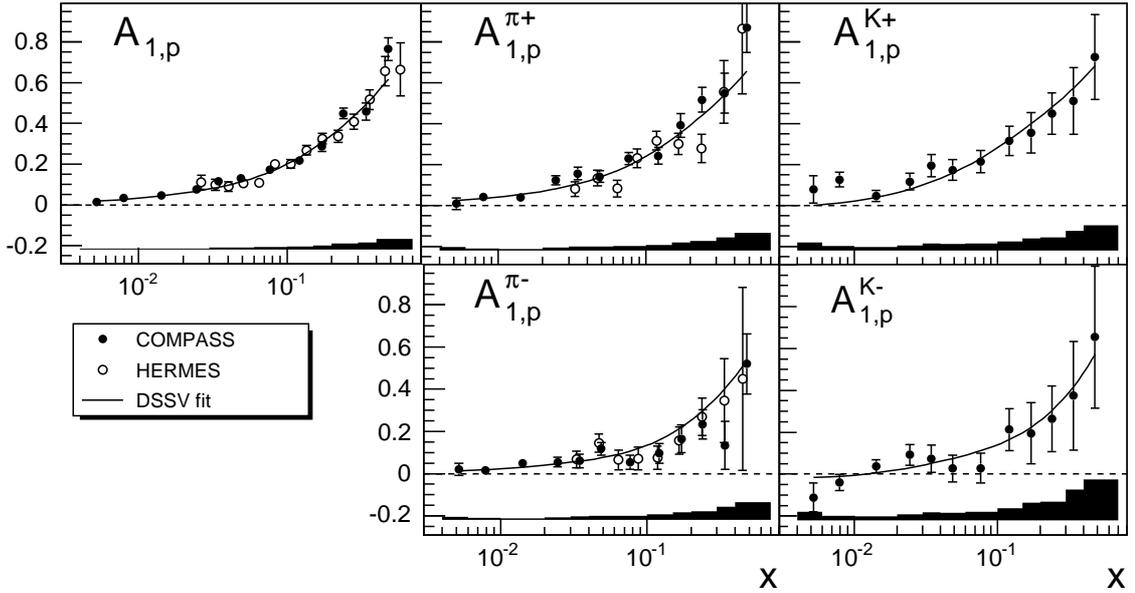}
\caption{The inclusive asymmetry $A_{1,p}$ \cite{g1p} and 
the semi-inclusive asymmetries $A_{1,p}^{\pi+}$, $A_{1,p}^{K+}$,
 $A_{1,p}^{\pi-}$,   $A_{1,p}^{K-}$ from the present 
measurements  
(closed circles). 
The bands  at the bottom of each plot show the
systematic errors.
The $A_{1,p}$, $A_{1,p}^{\pi+}$ and  $A_{1,p}^{\pi-}$ measurements 
from HERMES \cite{hrm1,hrm2} (open circles) are shown for 
comparison. The curves show the 
predictions of the DSSV fit \cite{dssv}.
}
\label{fig:asym_p}
\end{figure}

The semi-inclusive virtual-photon asymmetries for scattering of 
muons off a proton target
are listed in Table~\ref{tab:SIDIS_asym} with
their statistical and systematic errors. Since inclusive and 
semi-inclusive
asymmetries originate from the same events, their errors are 
correlated (see Table~\ref{tab:SIDIS_correl}).
The largest correlations ($\approx 0.4$) are those between 
inclusive and semi-inclusive pion
asymmetries, due to the larger pion multiplicity. The unfolding 
procedure also
generates a negative correlation between pions and kaons of the 
same charge. This correlation is
larger at small $x$ ($\approx -0.16$) due to the lower purity of 
the kaon sample.

The systematic errors contain a multiplicative part resulting from 
uncertainties of the
beam and target polarisations,  the dilution factor and  the 
ratio $R = \sigma^L/\sigma^T$
used to calculate the depolarisation factor\cite{e143}.
Added in quadrature, these uncertainties
amount to 6\% of the value of the asymmetry.  
An important contribution to the systematic
error arises from possible false asymmetries generated by 
instabilities in
the experimental setup. The effect of these random instabilities 
was evaluated by a 
statistical test on the asymmetries made on 23 subsets of data. 
At the level
of one standard deviation the upper bound of the error due to 
these time-dependent effects
is found to be $0.56\sigma_{stat}$.

The experimental double-spin asymmetries for a proton target are 
shown in Fig.~\ref{fig:asym_p}. They are compared  to the 
predictions 
of the DSSV fit \cite{dssv} at the $(x,Q^2)$ values of the data. 
The HERMES inclusive\cite{hrm2} and semi-inclusive\cite{hrm1} 
measurements for $\pi^+$ and $\pi^-$ are also displayed. The 
agreement  
with the DSSV parameterisation is good, even  for the kaon 
asymmetries for which no data were available when the prediction 
was made.
In spite of the different kinematic conditions, the agreement 
between the COMPASS and the HERMES values for the pion asymmetries 
is also good. This observation illustrates the fact that the
$Q^2$ dependence at fixed $x$ is  small for semi-inclusive 
asymmetries.

The spin asymmetries for a deuteron target were evaluated from 
our  previous
data obtained with a $^6$LiD target. The published values 
\cite{flavsep}  were corrected to account for the admixtures of  
$^7$Li (4.4\%) and  $^1$H (0.5\%) in the target
material\cite{neliba}. These isotopes are both polarised to more 
than 90\% \cite{ball}. The resulting
corrections, which do not exceed one fourth of the statistical 
error, are listed in Table~\ref{tab:SIDIS_correct} for each 
asymmetry and each bin of $x$.\footnote{These corrections should 
always be applied when using the data of Ref.~\cite{flavsep}.}
A similar correction to the inclusive
asymmetry $A_{1,d}$ has been used in Ref.~\cite{g1p}.

\section{Polarised PDFs from a LO fit to the asymmetries}

At LO in QCD and under the assumption of independent quark fragmentation,
the spin asymmetry for hadrons $h$ produced in the current fragmentation
region can be written as a sum of products of the quark, 
$\Delta q$,   and antiquark, $\Delta {\overline q}$, helicity
distributions with the corresponding fragmentation functions 
$D_q^h$ and $D_{\overline q}^h$:
\begin{eqnarray}
A_1^h(x,z) &=& \frac{\sum_q e_q^2 \Bigl(\Delta q(x) D_q^h(z) 
+ \Delta {\overline q(x)} D_{\overline q}^h(z)\Bigr)}
{\sum_q e_q^2 \Bigl(q(x) D_q^h(z) + {\overline q(x)} 
D_{\overline q}^h(z)\Bigr)} \;.
\label{indep_frag}
\end{eqnarray}

In the present analysis the $Q^2$ dependence of the asymmetries 
is neglected and all measurements are assumed to be valid at 
$Q_0^2 = 3$ (GeV/$c$)$^2$.  The LO unpolarised parton 
distributions (PDFs) with three quark flavours from the MRST parameterisation \cite{mrst} are used. The fragmentation
functions are taken from the LO parameterisation of 
DSS \cite{dss}. 
As in previous analyses \cite{smc,flavsep}, 
the unpolarised PDFs which are extracted from cross sections 
assuming
non-zero values of $R$ are corrected by a factor 
$1 + R(x,Q_0^2)$\cite{e143} to take into account the fact 
that $R$ is assumed to be zero at LO. 
The asymmetries for a deuteron target 
are corrected by the factor (1 -- $1.5\omega_D$) 
where $\omega_D$ is
the probability for a deuteron to be in a D-state 
($\omega_D = 0.05 \pm 0.01$) \cite{machl}. 
The four semi-inclusive asymmetries for a
proton target, the four semi-inclusive asymmetries for a 
deuteron target and the two inclusive asymmetries thus 
provide a system of ten
equations with six unknowns ($\Delta u$, $\Delta d$, 
$\Delta {\overline u}$, $\Delta {\overline d}$,
$\Delta s$ and $\Delta {\overline s}$).  A least-square fit 
to the data is performed independently in each 
bin of $x$. The analysis is limited to $x \le 0.3$ because the 
antiquark distributions become
insignificant above this limit.
Above $x = 0.3$, $\Delta u(x)$ and $\Delta d(x)$ are obtained 
from the inclusive structure functions $g_1^p(x)$ and $g_1^d(x)$ 
by assuming $\Delta {\overline q}$ to be zero. 
\begin{figure}[tb]
\centering
\includegraphics[width=0.75\textwidth,clip]{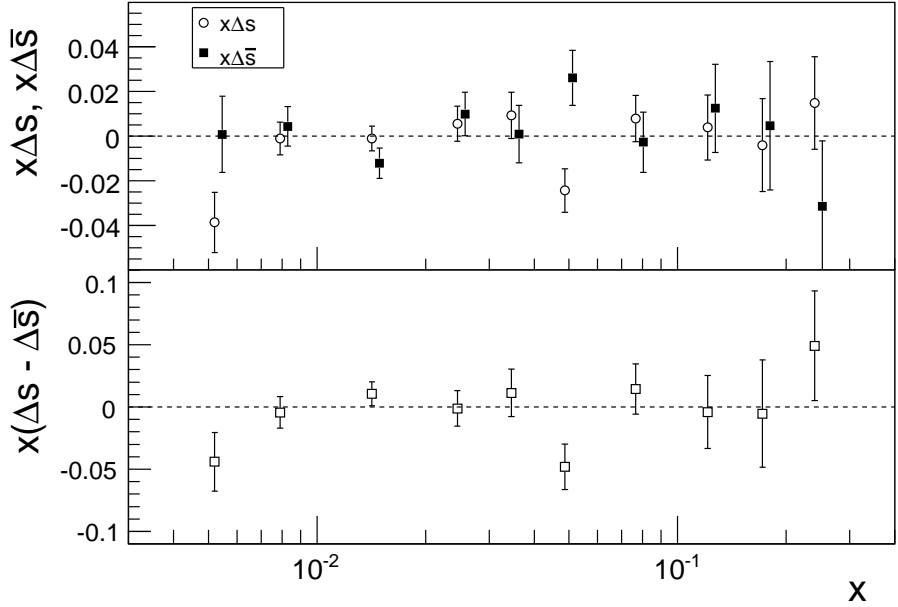}
\caption{Comparison of $x\,\Delta s$ (open circles) and 
$x\,\Delta {\overline s}$ (squares)
 at $Q_0^2 = 3$ (GeV/$c$)$^2$ (top) and corresponding values 
of the difference $x (\Delta s -  \Delta {\overline s})$ (bottom).
}
\label{fig:ds_dsbar}
\end{figure}

The fit results for the $\Delta s$ and $\Delta {\overline s}$ distributions and for their difference are displayed in 
Fig.~\ref{fig:ds_dsbar}. In the measured $x$ range both 
distributions are flat and compatible with zero. The same 
observation can be made for their difference, 
$\Delta s - \Delta {\overline s}$; 
 only  one point out of ten is outside  two standard
deviations  (2.7 $\sigma$ at $x = 0.0487$).
We have checked that the vanishing values of 
$\Delta s - \Delta {\overline s}$
are not artificially generated by the MRST parameterisation 
of the unpolarised PDFs where $s(x) = {\overline s}(x)$ is 
assumed.
The $s(x)$ and ${\overline s}(x)$ distributions were scaled simultaneously
by factors 2 and 0.5 and allowed to differ in any interval of 
$x$ 
by a factor of 2.
The values of $\Delta s(x)$ and $\Delta {\overline s(x)}$ were 
found to be nearly independent of these modifications.
We conclude that there is no significant difference between 
$\Delta s(x)$ and $\Delta {\overline s(x)}$ in the $x$-range 
covered by the data.
This conclusion remains valid when the DSS fragmentation 
functions used in the fit
are replaced by those derived by EMC \cite{emc}.
 The results on $\Delta s(x)$ and $\Delta {\overline s(x)}$ 
are at variance with
the SU(3) Chiral Quark-Soliton model prediction 
$ |\Delta s(x)| \gg |\Delta {\overline s(x)}|$
\cite{waka_1} but are compatible with statistical models  
predicting that $\Delta s(x) - \Delta {\overline s(x)}$
should be zero \cite{bhalerao} or small \cite{BSS}. 

\begin{figure}[tb]
\centering
\includegraphics[width=0.95\textwidth,clip]{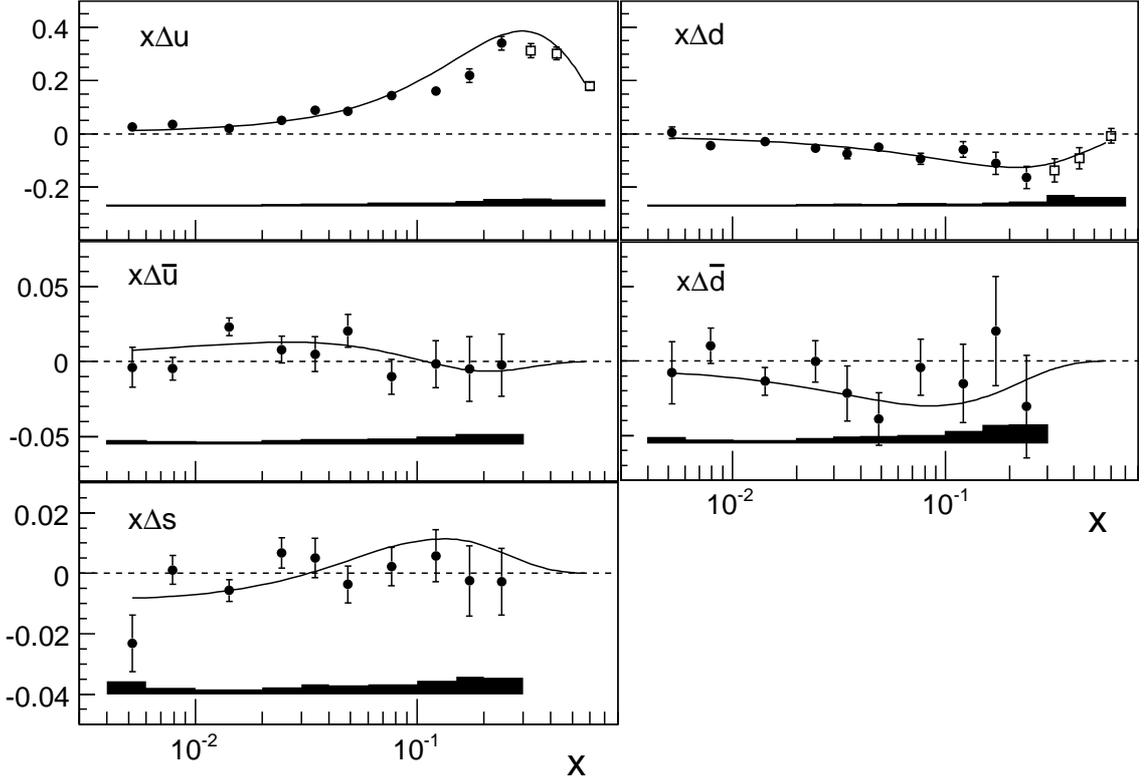}
\caption{The quark helicity distributions $x\,\Delta u$, 
$x\,\Delta d$, $x\,\Delta {\overline u}$,
$x\,\Delta {\overline d}$ and 
$x\,\Delta s$ at $Q_0^2 = 3$ (GeV/$c$)$^2$
as a function of $x$. The values for $x < 0.3$ (black dots)
are derived at LO from the COMPASS spin
asymmetries  using the DSS fragmentation functions
\cite{dss}. Those at $x > 0.3$ (open squares) are derived 
from
the values of the polarised structure function $g_1(x)$ 
quoted in \cite{g1p,g1d} assuming $\Delta {\overline q}
= 0$.
The bands at the bottom of each plot show the systematic 
errors.
The curves show the predictions of the DSSV fit calculated 
at NLO\cite{dssv}.
}
\label{fig:pol_pdf}
\end{figure}

From here on the distributions of $\Delta s$ and $\Delta 
{\overline s}$ will be assumed
to be equal, an assumption
which reduces the number of unknowns to five and improves the statistical
precision  of the fit results at least by a factor 1.5.
The $\chi^2$ of the fits varies from 1.8 to 8.5 in the different 
$x$ bins with an
average of 4.0 for 5 degrees of freedom, corresponding to a probability of 55\%.
Within their statistical precision the data are thus compatible 
with the
factorisation formula of Eq.~(\ref{indep_frag}).

The results for the quark helicity distributions 
$\Delta u $, $\Delta d $,
$\Delta {\overline u}$, $\Delta {\overline d}$ and $\Delta s$ ($\Delta s$ = $\Delta {\overline s}$) 
are shown in Fig.~\ref{fig:pol_pdf}. As for the asymmetries, they 
are in good qualitative agreement with the results from 
HERMES\cite{hrm1}. A quantitative comparison is not made here, 
since the HERMES helicity distributions are extracted under 
different assumptions for the fragmentation functions and for 
the unpolarised flavour distributions. In the range 
$0.3 < x < 0.7$ three additional values of $\Delta u$ and 
$\Delta d$, derived from the $g_1^p(x)$ and 
$g_1^d(x)$\cite{g1d} structure functions, are also displayed. 
The $g_1^d(x)$ values  include the target material 
corrections quoted in \cite{g1p}. 
The dominant contribution to the systematic error 
of $\Delta u$ and $\Delta d$
comes from the
uncertainty of the beam polarisation, which affects all data 
in the same way and leads to an uncertainty of
5\% for all fitted values. The systematic error on the 
antiquark and strange quark distributions 
is mainly due
to possible false asymmetries generated by time-dependent 
effects on the detector acceptance.
The curves show the results of the DSSV fit at Next-to-Leading 
Order (NLO) \cite{dssv}.
The comparison with 
the experimental results derived at LO is thus only  
qualitative. Nevertheless, the curves reproduce fairly
well the shape of the data, confirming a previous observation 
that a direct extraction at LO
provides  a good estimate of the shape of the helicity 
distributions \cite{oleg}. 
The antiquark distributions,
 $\Delta {\overline u}$ and $\Delta {\overline d}$, do not 
show any significant variation in the $x$ range
of the data, the former being consistent with zero, the latter 
being slightly negative. 

The values of the strange quark helicity distribution confirm 
with slightly reduced errors the results obtained from the 
deuteron data\cite{flavsep} alone.
With the same fragmentation functions (DSS)
no significant variation of $\Delta s(x)$ is observed in the 
range of the data. Only the first point at low $x$
shows a small deviation from zero ($\approx  2.5 \sigma$). 
This distribution is of special interest
due to the apparent contradiction between the SIDIS results 
and the  negative 
first moment derived \cite{g1d} from the spin structure function
$g_1(x)$.
The DSSV fit includes a negative contribution to $\Delta s$ for 
$x \leq 0.03$, which reconciles the inclusive and semi-inclusive results.
The evaluation of the first moment of $\Delta s(x)$ from  
inclusive measurements relies on the
value of the octet axial charge $a_8$, which is  derived from 
hyperon weak 
decays under the assumption of SU(3)$_{\rm f}$ symmetry.
A recent model calculation suggests that
$a_8$ may be substantially reduced and become close to the 
singlet axial
charge $a_0$ extracted from the data \cite{bass}. In this case 
the inclusive data would no longer imply
a negative value of $\Delta s$. Finally, as pointed out in our previous paper \cite{flavsep}, one
has to keep in mind that the semi-inclusive results on 
$\Delta s(x)$ strongly depend on
the choice of a set of fragmentation functions. This dependence 
is quantified in the next section.  

The first moments of the helicity distributions truncated to 
the range of the 
measurements are listed in Table~\ref{tab:SIDIS_truncated}.  
The missing contributions at low and at high $x$ have been 
evaluated by extrapolating the measured values and alternatively 
by using the DSSV parameterisation \cite{dssv}. The contributions 
at high $x$ are all small and do not exceed 0.01. The two 
methods lead to similar values for the valence quark moments 
$\Delta u_v = \Delta u - 
\Delta {\overline u}$ and $\Delta d_v = \Delta d  - \Delta 
{\overline d}$. In contrast, they differ for the sea quark 
moments and particularly for $\Delta s$ due to the 
sizable low-$x$ contribution assumed in the DSSV fit. The 
resulting full first moments for both methods are listed in 
Table~\ref{tab:SIDIS_full}.  The sum of  the quark and 
antiquark contributions 
$\Delta \Sigma = 0.32 \pm 0.03(stat.)$, obtained by linearly 
extrapolating the data,
is nearly identical to the value of 
$a_0=0.33 \pm 0.03(stat.)$\footnote 
{The admixture of $^7$Li and $^1$H in the target material
reduces the value of $a_0$ quoted in Ref.~\cite{g1d} by 
0.02\cite{g1p}.}  derived \cite{g1d} from the 
first moment of $g_1^d(x)$ using the octet 
axial charge $a_8$.   
Not surprisingly, the extrapolation 
with the DSSV parameterisation results in a much smaller 
value 
for $\Delta \Sigma$. The observed difference comes mainly 
from the negative behaviour of $\Delta s$ assumed at small 
$x$. The sum of the valence quark contributions 
$\Delta u_v + \Delta d_v = 0.39 \pm 0.03(stat.)$ 
is also consistent with our previous
determination based on the difference asymmetry of positive 
and negative hadrons in 
a subsample of the present deuteron 
data $(0.41 \pm 0.07(stat.)$ at $Q_0^2 = 10~(\GeV/c)^2)$\cite{val}. 

\begin{figure}[t]
\centering
\includegraphics[width=0.95\textwidth,clip]{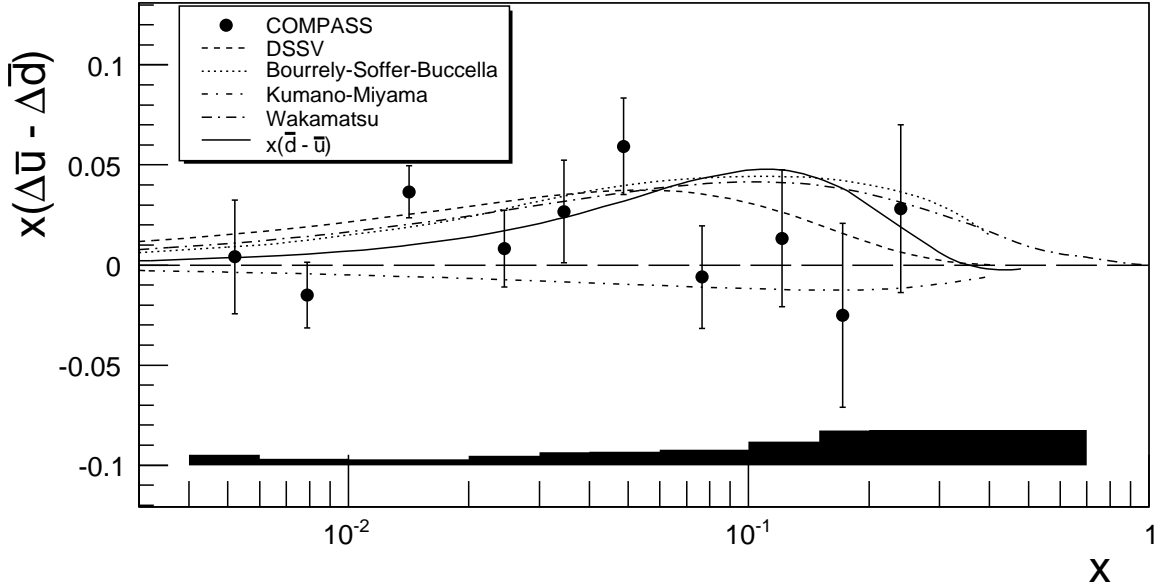}
\caption{The  flavour  asymmetry of the helicity distribution of 
the sea 
$x \,(\Delta {\overline u} - 
\Delta {\overline d})$ at $Q_0^2 = 3~(\GeV/c)^2$. 
The shaded area displays the systematic error. The 
dashed curve is the result of the DSSV fit at NLO. The other 
curves are model predictions from Wakamatsu\cite{waka_1} (long 
dash-dotted line), Kumano and Miyama \cite{kuma} (short 
dash-dotted line) and Bourrely, Soffer and Buccella \cite{bsb} (dotted line). 
The solid curve shows the MRST parameterisation 
for the unpolarised difference 
$x\,({\overline d} - {\overline u})$ at NLO.    
}
\label{fig:flav-asym}
\end{figure}

The   flavour  asymmetry of the helicity 
distribution  of the sea, 
$\Delta {\overline u} - \Delta {\overline d}$, is shown in 
Fig.~\ref{fig:flav-asym}. Although compatible with zero, 
the values indicate a slightly positive distribution. The 
DSSV fit at NLO\cite{dssv} and the unpolarised asymmetry 
${\overline d} - {\overline u}$ are shown for comparison. 
The first moment $\Delta {\overline u} - \Delta {\overline d}$ truncated to the range $0.004 < x < 0.3$ is  $0.06 \pm 0.04 
({\rm stat.}) \pm 0.02 ({\rm syst.})$. It is worth noting that 
the polarised first moment is about one 
standard deviation smaller than the unpolarised one 
truncated to the same range ($\approx 0.10$ for the
MRST parameterisation\cite{mrst}). The data thus disfavour 
models
predicting 
$\Delta {\overline u} - \Delta {\overline d} 
\gg {\overline d} - {\overline u}$
(see Refs.~\cite{peng,waka} and references therein). Three 
model predictions are shown in Fig.~\ref{fig:flav-asym}. The statistical model 
of Ref.~\cite{bsb} and the SU(3) version of the Chiral 
Quark--Soliton model of Ref.~\cite{waka_1} both predict 
positive distributions, while the Meson Cloud model of 
Ref.~\cite{kuma} predicts a slightly negative distribution. 
Within the statistical errors, the COMPASS data are compatible 
with all three predictions. 
The sum of the light quark helicity distributions, 
$\Delta {\overline u} + \Delta {\overline d}$, 
is mainly constrained by the deuteron data and nearly 
identical to the result published 
in Ref.~\cite{flavsep}. The first moment truncated to the 
range of the data is found
to be $-0.03 \pm 0.03 ({\rm stat.}) \pm 0.01 ({\rm syst.})$. 

\section{Influence of the fragmentation functions on the helicity 
distributions}

\begin{figure}[t]
\centering
\includegraphics[width=0.95\textwidth,clip]{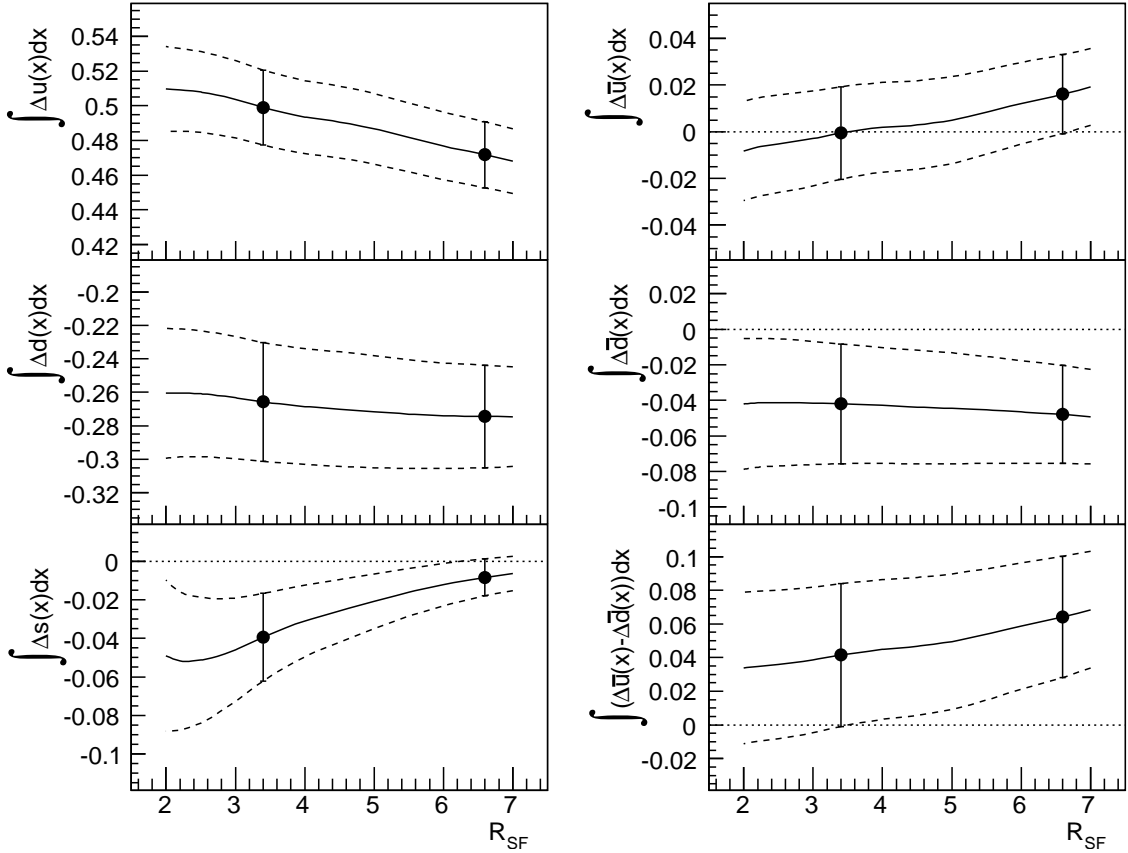}
\caption{Variation of the quark first moments 
$\Delta u$, $\Delta {\overline u}$, 
$\Delta d$, $\Delta {\overline d}$,
$\Delta s$ and
$\Delta {\overline u} - \Delta {\overline d}$ integrated over
the interval $0.004 < x < 0.3$ as a function of the ratio 
$R_{SF}$ of ${\overline s}$ and $u$ quark fragmentation functions 
into $K^+$. The ratio $R_{UF}$ is varied linearly from 0.13 at 
$R_{SF} = 6.6$ to 0.35 at $R_{SF} = 3.4$. The left and right black 
points indicate the values obtained using the EMC
\cite{emc} and the DSS \cite{dss} kaon fragmentation functions, 
respectively. 
}
\label{fig:scan_combi}
\end{figure}

The relation between the semi-inclusive 
asymmetries and the quark helicity distributions (Eq.~(\ref{indep_frag})) depends 
only on the ratios of fragmentation functions integrated over the 
selected range of $z$ $(0.2 < z < 0.85)$. Relevant for 
the kaon asymmetries are the unfavoured-to-favoured FF ratio, 
$R_{UF}$,  and strange-to-favoured FF ratio, $R_{SF}$: 
\begin{eqnarray}
R_{UF} = \frac{ \int D_d^{K^+}(z) \id z}{\int D_u^{K^+}(z) \id z},
~~~
R_{SF} = \frac{ \int D_{\overline s}^{K^+}(z) \id z}
{\int D_u^{K^+}(z) \id z}.
\label{d321}
\end{eqnarray}
In the DSS parameterisation, the $R_{UF}$ and $R_{SF}$ ratios are 
equal to 0.13 and 6.6 respectively. In the earlier EMC 
parameterisation\cite{emc} $R_{SF}$ is substantially smaller, 
$R_{SF} =  3.4$, whereas $R_{UF}$ is larger, $R_{UF} = 0.35$. 
Since the pion fragmentation functions are better
constrained by the data than the kaon ones, the effect of the 
corresponding ratios on the final result is expected to be much 
smaller. 
The dependence of the truncated moments quoted in 
Table~\ref{tab:SIDIS_truncated}  was then evaluated by 
varying $R_{SF}$ from 
$R_{SF} = 2.0$ to $R_{SF} = 7.0$. In order to keep the
$K^+$ multiplicity approximately constant, the value of $R_{UF}$ 
was simultaneously varied from 0.45 to 0.10 according to the 
relation $R_{UF} = 0.35 - 0.07 (R_{SF} -3.4)$.  
The resulting truncated first moments $\Delta u$, $\Delta 
{\overline u}$,
$\Delta d$, $\Delta {\overline d}$,
$\Delta s$ and $\Delta {\overline u} - \Delta {\overline d}$ are 
shown in Fig.~\ref{fig:scan_combi} as a function of  $R_{SF}$. 
We observe that the values of $\Delta u$ ($\Delta {\overline u}$)
increase (decrease) by more than one standard deviation when the 
ratios evolve
from the DSS to the EMC values. In contrast both $\Delta d$ and 
$\Delta {\overline d}$ remain nearly constant.
The variation of $\Delta s$ is much more pronounced: its value 
evolves from $-0.01$ to $-0.04$,
although with a much larger error. 
The difference $\Delta {\overline u} - \Delta {\overline d}$
follows the same trend as $\Delta {\overline u}$. It slightly 
decreases with $R_{SF}$, down to one standard deviation from 
zero at $R_{SF}=3.4$.  
We note that the simultaneous changes of the two ratios, while 
leaving the $K^+$ rate
practically unchanged, affect the $K^-$ rate only for $x \ge 0.1$. 
Precise values of $R_{UF}$ and $R_{SF}$ may thus be difficult 
to extract from the data. 

\section {Conclusions}

Longitudinal spin asymmetries for identified charged pions and 
kaons in semi-inclusive muon scattering on a proton target have 
been measured.
% by the COMPASS Collaboration at CERN. 
%The data  
%significantly improve the statistical accuracy of the pion 
%asymmetries and provide the first measurements of the kaon 
%asymmetries. 
The pion data extend the 
measured region by an order of magnitude towards small $x$, while
the kaon asymmetries for the proton were measured for the first 
time.
The new SIDIS asymmetries for the proton were combined with our 
previous SIDIS asymmetries for the deuteron and with both proton 
and deuteron inclusive measurements in order 
to evaluate the three lightest flavour quark and antiquark 
helicity distributions. The resulting $\Delta u$ and
$\Delta d$  distributions are dominant at medium and high $x$. 
The values of the antiquark distributions are small and do not 
show any significant variation in the measured range. The 
$\Delta {\overline u}$ distribution is consistent with zero, 
while $\Delta {\overline d}$ seems to indicate a slightly 
negative behaviour. Accordingly, the flavour asymmetry of the 
helicity distribution of the sea,
$\Delta {\overline u} - \Delta {\overline d}$, is slightly 
positive, about 1.5 standard deviations from zero. No difference 
is observed between the $\Delta s$ and $\Delta {\overline s}$ 
distributions, which are both 
compatible with zero over the measured $x$ range. The sum of the 
flavour-separated first moments, linearly extrapolated to $x=0$, 
is in good agreement with our previous determination of 
$\Delta \Sigma$ based on the first moments of the spin structure 
function $g_1^d(x)$. 
The dependence of the results on the fragmentation functions used 
was evaluated. Sizable for $\Delta u$ and $\Delta {\overline u}$ 
distributions, this dependence becomes critical for the $\Delta s$ 
distribution.  
\section*{Acknowledgements}

We gratefully acknowledge the support of the CERN management 
and staff and
the skill and effort of the technicians of our collaborating 
institutes.
Special thanks go to V.~Anosov and V.~Pesaro for their 
technical support
during the installation and the running of this experiment.
%This work was made possible by the financial support of our 
%funding agencies.
~

%\section{Tables}
\begin{sidewaystable}[here]
\caption{Unfolded  asymmetries for charged pions and kaons produced 
on a proton target. The first error is statistical, 
the second is systematic.}
\vspace{0.3cm}
\centering
{%\small
%\hspace{-1.2cm}
%\begin{tabular}{|c|r|r|r|r|r|}
\begin{tabular}{crrrrr}
\hline
\hline
$\langle x\rangle$ &$\langle Q^2\rangle$~~~~ &
$A_{1,p}^{\pi+}$~~~~~~~~~~~~~~ & $A_{1,p}^{\pi-}$~~~~~~~~~~~~~~  & 
$A_{1,p}^{K+}$~~~~~~~~~~~~~~  & $A_{1,p}^{K-\strut}$~~~~~~~~~~~~~~  \\
                   &   (GeV/$c)^2$    &      &   &   &  \\
\hline
0.0052 &  1.16~~~~ & $  0.008\pm0.029\pm0.016$ & 
$  0.020\pm0.029\pm0.016$ & $  0.078\pm0.067\pm0.038$ & 
$ -0.112\pm0.069\pm0.039$ \\
0.0079 &  1.46~~~~ & $  0.041\pm0.018\pm0.010$ & 
$  0.016\pm0.018\pm0.010$ & $  0.126\pm0.036\pm0.021$ & 
$ -0.040\pm0.039\pm0.022$ \\
0.0142 &  2.12~~~~ & $  0.040\pm0.014\pm0.008$ & 
$  0.049\pm0.015\pm0.009$ & $  0.046\pm0.028\pm0.016$ & 
$  0.038\pm0.031\pm0.018$ \\
0.0245 &  3.22~~~~ & $  0.122\pm0.022\pm0.014$ & 
$  0.055\pm0.023\pm0.013$ & $  0.117\pm0.041\pm0.024$ & 
$  0.092\pm0.048\pm0.028$ \\
0.0346 &  4.36~~~~ & $  0.156\pm0.030\pm0.019$ & 
$  0.060\pm0.032\pm0.018$ & $  0.196\pm0.054\pm0.033$ & 
$  0.074\pm0.066\pm0.037$ \\
0.0487 &  5.97~~~~ & $  0.141\pm0.029\pm0.018$ & 
$  0.118\pm0.031\pm0.019$ & $  0.174\pm0.051\pm0.031$ & 
$  0.027\pm0.064\pm0.036$ \\
0.0765 &  8.96~~~~ & $  0.230\pm0.031\pm0.022$ & 
$  0.053\pm0.033\pm0.019$ & $  0.215\pm0.054\pm0.033$ & 
$  0.029\pm0.071\pm0.040$ \\
0.121~ & 13.8~~~~~ & $  0.243\pm0.041\pm0.027$ & 
$  0.096\pm0.047\pm0.027$ & $  0.315\pm0.072\pm0.044$ & 
$  0.212\pm0.101\pm0.058$ \\
0.172~ & 19.6~~~~~ & $  0.392\pm0.058\pm0.040$ & 
$  0.165\pm0.066\pm0.038$ & $  0.355\pm0.099\pm0.059$ & 
$  0.195\pm0.147\pm0.083$ \\
0.240~ & 27.6~~~~~ & $  0.518\pm0.060\pm0.046$ & 
$  0.233\pm0.069\pm0.041$ & $  0.450\pm0.101\pm0.063$ & 
$  0.264\pm0.157\pm0.089$ \\
0.341~ & 40.1~~~~~ & $  0.549\pm0.097\pm0.064$ & 
$  0.134\pm0.113\pm0.064$ & $  0.512\pm0.163\pm0.097$ & 
$  0.375\pm0.259\pm0.147$ \\
0.480~ & 55.6~~~~~ & $  0.871\pm0.122\pm0.086$ & 
$  0.520\pm0.142\pm0.085$ & $  0.726\pm0.207\pm0.124$ & 
$  0.654\pm0.339\pm0.194$ \\ 
\hline
\hline
\end{tabular}
}
\label{tab:SIDIS_asym}
\end{sidewaystable}

\clearpage
\begin{sidewaystable}[here]
\caption{Correlation coefficients $\rho$ of the unfolded asymmetries 
in bins of $x$.}
\vspace{0.3cm}
\label{tab:corr}

{%\tiny
%\begin{tabular}{|c||r|r|r|r|r|r|r|r|r|r|r|r|}
\begin{tabular}{crrrrrrrrrrrr}
\hline
\hline
$x$-bin & $\!\!$0.004--0.006$\!\!$ &$\!\!$0.006--0.01$\!\!$ &
$\!$0.01--0.02$\!$ &$\!$0.02--0.03$\!$ &$\!$0.03--0.04$\!$ &
$\!$0.04--0.06$\!$ &$\!$0.06--0.10$\!$ &$\!$0.10--0.15$\!$ &
$\!$0.15--0.20$\!$ &$\!$0.2--0.3$\!$ &$\!$0.3--0.4$\!
$ &$\!$0.4--0.7$\!$\\
\hline
$\rho(A_{1,p}^{\pi+},A_{1,p})$        & 0.29 & 0.34 & 0.37 &  0.38 
&  0.39 &  0.40 &  0.41 &  0.42 &  0.43 &  0.43 & 0.45 & 0.46\\
\hline
$\rho(A_{1,p}^{\pi-},A_{1,p})$        &  0.30 &  0.34 &  0.37 &  0.38 
&  0.38 &  0.39 &  0.39 &  0.39 &  0.38 &  0.38 & 0.39 & 0.40\\
$\rho(A_{1,p}^{\pi-},A_{1,p}^{\pi+})$ &  0.12 &  0.15 &  0.17 &  0.16 
&  0.15 &  0.16 &  0.16 &  0.15 &  0.16 &  0.16 & 0.19 & 0.20\\
\hline
  $\rho(A_{1,p}^{K+},A_{1,p})$        &  0.26 &  0.28 &  0.28 &  0.26 
&  0.27 &  0.28 &  0.29 &  0.30 &  0.29 &  0.29 & 0.30 & 0.30\\
  $\rho(A_{1,p}^{K+},A_{1,p}^{\pi+})$ & $-0.17$ & $-0.09$ & $-0.04$ & 
$-0.02$ & $-0.02$ & $-0.01$ & $-0.02$ & $-0.01$ & $-0.01$ & $-0.02$ & 
$-0.02$ & $-0.01$\\
  $\rho(A_{1,p}^{K+},A_{1,p}^{\pi-})$ &  0.03 &  0.04 &  0.04 &  0.05 
&  0.05 &  0.05 &  0.05 &  0.06 &  0.05 &  0.05 & 0.04 & 0.03\\
\hline
  $\rho(A_{1,p}^{K-},A_{1,p})$        &  0.12 &  0.15 &  0.17 &  0.17 
&  0.17 &  0.18 &  0.17 &  0.17 &  0.16 &  0.16 & 0.16 & 0.15\\
  $\rho(A_{1,p}^{K-},A_{1,p}^{\pi+})$ &  0.03 &  0.03 &  0.04 &  0.04 
&  0.04 &  0.04 &  0.03 &  0.04 &  0.02 &  0.03 & 0.02 & 0.05\\
  $\rho(A_{1,p}^{K-},A_{1,p}^{\pi-})$ & $-0.16$ & $-0.09$ & $-0.05$ & 
$-0.03$ & $-0.03$ & $-0.03$ & $-0.03$ & $-0.03$ & $-0.02$ & $-0.03$ & 
$-0.04$ & $-0.02$\\
  $\rho(A_{1,p}^{K-},A_{1,p}^{K+})$   &  0.05 &  0.08 &  0.10 &  0.10 
&  0.10 &  0.11 &  0.11 &  0.12 &  0.11 &  0.11 & 0.13 & 0.16 \\
\hline
\hline
\end{tabular}
}
\label{tab:SIDIS_correl}
\end{sidewaystable}

\begin{table}[here]
\caption{\small Corrections to  the deuteron spin asymmetries 
$A_{1,d}^h$ and
$A_{1,d}$ due to admixture of $^7$Li and $^1$H into the $^6$LiD
target material. The corrections are to be subtracted from the 
values of Ref.~\cite{flavsep}.}
\vspace{0.3cm}
\centering
%\begin{tabular}{|c|l|l|l|l|l|}
\begin{tabular}{clllll}
\hline  \hline
$x$ range  & $\pi^+$ & $\pi^-$  & $K^+$ & $K^{-\strut}$  & incl. \\
\hline
0.004--0.006 &  0.001&  0.~~~&  0.001&  0.~~~&  0.~~~ \\
0.006--0.010 &  0.001&  0.~~~&  0.001&  0.~~~&  0.001 \\
0.010--0.020 &  0.001&  0.001&  0.002&  0.~~~&  0.001 \\
0.020--0.030 &  0.002&  0.001&  0.002&  0.001&  0.001 \\
0.030--0.040 &  0.002&  0.001&  0.003&  0.001&  0.002 \\
0.040--0.060 &  0.003&  0.001&  0.003&  0.001&  0.002 \\
0.060--0.100 &  0.004&  0.002&  0.005&  0.002&  0.003 \\
0.100--0.150 &  0.006&  0.002&  0.006&  0.003&  0.004 \\
0.150--0.200 &  0.008&  0.003&  0.008&  0.004&  0.006 \\
0.200--0.300 &  0.011&  0.004&  0.010&  0.005&  0.008 \\
0.300--0.400 &  0.015&  0.005&  0.013&  0.009&  0.011 \\
0.400--0.700 &  0.020&  0.006&  0.017&  0.013&  0.015 \\
\hline  \hline
\end{tabular}
\label{tab:SIDIS_correct}
\end{table}

\begin{table}[here]
\caption{\small First moments of the quark helicity distributions
at $Q_0^2 = 3$ (GeV/$c$)$^2$ truncated to the range of the 
measurements
and derived with the DSS fragmentation functions.
The first error is statistical, the second one systematic.
The values of the sea quark distributions 
 for $x \ge 0.3$ are assumed to be zero.}
\vspace{0.3cm}
\centering
%\begin{tabular}{|c|r|r|}
\begin{tabular}{crr}
\hline  \hline
$x$ range  & $0.004 < x < 0.3$  & $ 0.004 < x < 0.7$  \\
\hline
$\Delta u$ & $0.47 $  $\pm$ 0.02  $\pm$ 0.03   & 
$0.69 $  $\pm$ 0.02  $\pm$ 0.03   \\
$\Delta d$ & $-0.27 $ $\pm$ 0.03  $\pm$ 0.02  & 
$-0.33 $ $\pm$ 0.04  $\pm$ 0.03   \\
$\Delta {\overline u}$ & 0.02  $\pm$ 0.02  
$\pm$ 0.01   & ---~~~~~~~~~~~~~~\\
$\Delta {\overline d}$ & $-0.05 $ $\pm$ 0.03  
$\pm$ 0.02  & ---~~~~~~~~~~~~~~\\
$\Delta s (\Delta {\overline s}) $  & 
$-0.01 $ $\pm$ 0.01  $\pm$ 0.01  & ---~~~~~~~~~~~~~~ \\
\hline  
$\Delta u_v$ &  0.46  $\pm$ 0.03  $\pm$ 0.03    
& 0.67  $\pm$ 0.03  $\pm$ 0.03   \\
$\Delta d_v$ & $-0.23 $ $\pm$ 0.05  $\pm$ 0.02  
& $-0.28 $ $\pm$ 0.06  $\pm$ 0.03   \\
$\Delta {\overline u}-\Delta {\overline d}$ 
& 0.06 $\pm$ 0.04 $\pm$ 0.02 & ---~~~~~~~~~~~~~~\\
$\Delta {\overline u}+\Delta {\overline d}$ 
& $-0.03$ $\pm$ 0.03 $\pm$ 0.01 & ---~~~~~~~~~~~~~~\\
$\Delta \Sigma$ & 0.15 $\pm$ 0.02 $\pm$ 0.02 
& 0.31 $\pm$ 0.03 $\pm$ 0.03  \\
\hline  \hline
\end{tabular}
\label{tab:SIDIS_truncated}
\end{table}

\begin{table}[here]
\caption{\small Full first moments of the quark helicity 
distributions at $Q_0^2 = 3$ (GeV/$c$)$^2$. The unmeasured 
contributions at low and high $x$ were estimated by extrapolating 
the  data towards $x=0$ and $x=1$ and by using 
the DSSV parameterisation\cite{dssv}}
\vspace{0.3cm}
\centering
%\begin{tabular}{|c|r|r|}
\begin{tabular}{crr}
\hline  \hline
     & {\rm Extrapolation}  & {\rm DSSV} \\
\hline
$\Delta u$ & $0.71 $  $\pm$ 0.02  $\pm$ 0.03   
& $0.71 $  $\pm$ 0.02  $\pm$ 0.03   \\
$\Delta d$ & $-0.34 $ $\pm$ 0.04  $\pm$ 0.03  
& $-0.35 $ $\pm$ 0.04  $\pm$ 0.03   \\
$\Delta {\overline u}$ & 0.02  $\pm$ 0.02  $\pm$ 0.01   
& 0.03 $\pm$ 0.02 $\pm$ 0.01\\
$\Delta {\overline d}$ & $-0.05 $ $\pm$ 0.03  $\pm$ 0.02  
& $-0.07$ $\pm$ 0.03 $\pm$ 0.02\\
$\Delta s (\Delta {\overline s}) $ & $-0.01 $ $\pm$ 0.01 $\pm$ 0.01  
& $-0.05$ $\pm$ 0.01 $\pm$ 0.01 \\
\hline  
$\Delta u_v$ &  0.68  $\pm$ 0.03  $\pm$ 0.03    
& 0.68  $\pm$ 0.03  $\pm$ 0.03  \\
$\Delta d_v$ & $-0.29 $ $\pm$ 0.06  $\pm$ 0.03  
& $-0.28 $ $\pm$ 0.06  $\pm$ 0.03   \\
$\Delta \Sigma$ & 0.32 $\pm$ 0.03 $\pm$ 0.03 
& 0.22 $\pm$ 0.03 $\pm$ 0.03  \\
\hline  \hline
\end{tabular}
\label{tab:SIDIS_full}
\end{table}
\end{document}